\documentstyle[twoside,fleqn,espcrc2,epsf]{article}
\newcommand{\noi}{\noindent}
\newcommand{\eq}{\begin{equation}}
\newcommand{\en}{\end{equation}}
\newcommand{\eqa}{\begin{eqnarray}}
\newcommand{\ena}{\end{eqnarray}}
\title{Zero-momentum modes and chiral limit in compact lattice QED
\thanks{Talk presented by N.V.~Zverev}
}

\author{
I.L.~Bogolubsky$^{\rm a}$, V.K.~Mitrjushkin%
\address{Joint Institute for Nuclear Research, 141980 Dubna, Russia},
M.~M\"uller-Preussker$^{\rm b}$ and N.V.~Zverev%
\address{Humboldt-Universit\"at zu Berlin, Institut f\"ur Physik,
Invalidenstr. 110, D-10115, Germany}
}
\begin{document}

\begin{abstract}
%-------------
The influence of zero-momentum gauge modes on
physical observables is investigated for compact
lattice QED with dynamical and quenched Wilson fermions. Within
the Coulomb phase, zero-momentum modes are shown to
hide the critical behavior of gauge invariant
fermion observables near the chiral limit. Methods
for eliminating zero-momentum modes are discussed.
\vspace{1pc}
\end{abstract}

\maketitle

Lattice gauge theories allow to compute physical observables without
gauge fixing. At the same time, more detailed information about
nonperturbative properties of quantum fields can be extracted from
gauge-dependent objects. But the straightforward application of
iterative gauge fixing procedures leads to the appearance of gauge 
(Gribov) copies. For compact lattice QED in the Coulomb phase, such spurious
gauge configurations disturb the correct perturbative behavior of
both the photon and fermion correlators \cite{NaSi,NaPl,BoMiMPPa}. Numerical
\cite{BoMiMPPa,BoMiMPPe,BoMiDD} and analytical \cite{Mitr} investigations
of this problem have shown the main excitations responsible for the occurence 
of such copies to be double Dirac sheets and zero-momentum modes.
The former can be removed by usual, or periodic, gauge rotations. However,
the removal of the latter requires constant, or nonperiodic, 
gauge transformations, which -- for finite lattice size -- 
violate the invariance of the fermion matrix.

In this talk we present results of a study of the influence of 
zero-momentum modes in compact lattice QED within the physical 
(Coulomb) phase.

The action $S$ is given by

\eqa %\label{eq1}
S &=& S_{G}[U] + S_{F}[U, {\bar \psi}, \psi]\,,
\\
\nonumber \\
S_{G} &=& \beta \sum_{x,\mu > \nu}%
\bigl( 1 - {\rm Re}\,U_{x\mu}U_{x+\hat{\mu},\nu}%
U^{\ast}_{x+\hat{\nu},\mu}U^{\ast}_{x\nu}\bigr),
\nonumber \\
\nonumber \\
S_{F} &=& \sum_{{\rm f}=1}^{N_{\rm f}}\sum_{x,y}
{\bar \psi}^{\rm f}_{x}{\cal M}_{xy} \psi^{\rm f}_{y},
\nonumber
\ena

\noi
where ${\cal M}$ is the Wilson fermion matrix

\eqa
{\cal M}_{xy} \equiv  {\bf 1_{\rm D}}\delta_{xy} - \kappa
\sum_{\mu}\Bigl\{( {\bf 1_{\rm D}} - \gamma_{\mu}) U_{x\mu}
        \delta_{y, x+\hat{\mu}}
\\ + ( {\bf 1_{\rm D}} + \gamma_{\mu})
        U_{y\mu}^{\ast}\delta_{y, x-\mu}\Bigr\}\,.
\nonumber
\ena

\noi
$U_{x\mu} = \exp ({\rm i}\theta_{x\mu})$,
$\theta_{x\mu} \in (-\pi, \pi]$ denote the link variables.
$\beta$ and $\kappa$ are the inverse square bare coup\-ling and the
hopping parameter, respectively. The gauge as well as the fermion
field obey periodic boundary conditions (b.c.) except for the $x_4$
(time) direction, for which the fermion field will be taken either 
periodic or antiperiodic.

In what follows we consider both the quenched case simulated
with a heat bath method and the dynamical case for
$N_{\rm f}=2$ flavors studied with the hybrid Monte-Carlo method (HMC).

When studying gauge-dependent objects like gauge link correlators or the
fermion propagator the Lorentz (or Landau) gauge fixing procedure is
usually applied by iteratively maximizing the functional

\eq
F[\Omega]
\equiv \frac{1}{4V}\sum_{x\mu} {\rm Re\,}U_{x\mu}^{\Omega}
\longrightarrow {\rm max}\,,
           \label{func_lg}
\en

\noi
with $U_{x\mu}^\Omega = \Omega_x U_{x\mu}\Omega^{\ast}_{x+\mu}$
and $\Omega_x \in {\rm U(1)}$. $V=N_s^3 \cdot N_4$ denotes the lattice volume.
The gauge functional in most cases has many local extrema. Consequently
the iterative maximization procedure can provide different gauge copies.
But the 'best' or physically correct gauge copies are expected to be related
to the global maxima.

%=======================================================================
%   FIGURE 1.
%   Pion norm in the Coulomb phase.
%-----------------------------------------------------------------------
\begin{figure}[tbp]
\vspace{1.2cm}
\hspace*{0.5cm}
\epsfxsize=6.5cm\epsfysize=8cm\epsffile{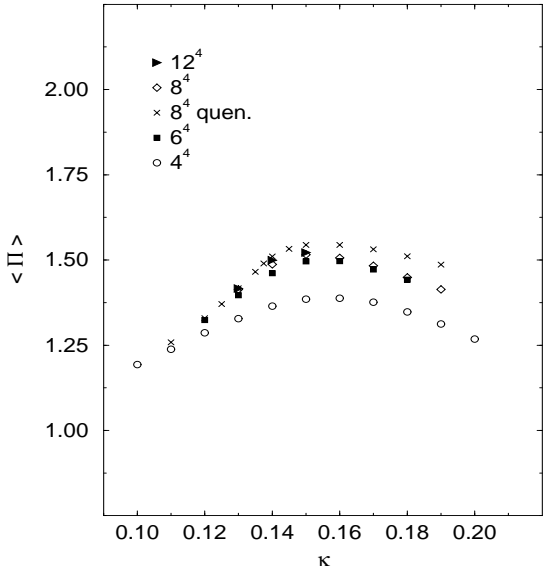}
\vspace{-1.0cm}
\caption{Pion norm as function of $\kappa$ for full (and quenched)
compact QED with Wilson action at $\beta=1.1$ for various lattice sizes
(data taken from \cite{HfMiMPNhSt}).
}
\label{fig.1}
\end{figure}
%=======================================================================

In our recent paper \cite{BoMiMPPeZv} gauge variant
fermion correlators in the quenched approximation have been calculated
within the standard Lorentz gauge.  It has been shown that these correlators
behave strongly decaying, as if they were massive in the limit
$\kappa\to\kappa_c(\beta)$.
This effect has been shown caused by constant or zero--momentum modes
of the gauge fields

\eq
z_\mu[U]\equiv \frac{1}{V} \sum_x \theta_{x\mu}\,.
             \label{zmm}
\en

\noi
Within the Lorentz gauge the zero--momentum modes do not allow to
obtain the best gauge copy, i.e.  the maximization of the functional 
$F[\Omega]$
defined in  eq. (\ref{func_lg}) does not lead to its absolute maximum.
Zero--momentum modes cannot be eliminated by usual (periodic)
gauge transformations. The Lorentz gauge can only drive them into the
interval $[-\pi/N_{\mu},+\pi/N_{\mu}]$ \cite{BoMiMPPeZv} which, however,
appears to be sufficient to distort gauge-dependent
observables. Increasing the lattice volume does not change the situation
significantly.

In \cite{BoMiMPPe} an iterative Lorentz gauge fixing procedure
combined with zero-momentum mode subtraction was proposed (ZML gauge).
The zero-momentum mode suppression

\eq
z_\mu[U]=0
\en

\noi
is achieved by appropriate non-periodic gauge transformation steps
\eq
U_{x\mu} \to  c_\mu U_{x\mu}, \qquad c_\mu \in {\rm U(1)}
                   \label{npt}
\en
following each Lorentz gauge step.

Within the ZML gauge the Lorentz functional (\ref{func_lg}) reaches
its global maximum with high accuracy \cite{BoMiMPPe,BoMiDD}. On the
other hand, ZML gauge fixing leads to the correct perturbative
behavior of gauge-dependent objects \cite{BoMiMPPe,BoMiMPPeZv}. In
particular, the results are compatible with vanishing fermion masses 
near the chiral limit.

In this talk, instead of gauge variant objects we consider the gauge
invariant -- with respect to periodic gauge transformations -- 'pion norm'

\begin{displaymath}
\langle \Pi \rangle = \left\langle \frac{1}{4V} {\rm Tr}
\left( {\cal M}^{-1}
\gamma_{5} {\cal M}^{-1}\gamma_{5} \right) \right\rangle
\sim \left\langle \sum_i \frac{1}{\lambda_i^2} \right\rangle,
\end{displaymath}

\noi
where $\langle \cdots \rangle$ means the functional average with
respect to the compact $U(1)$ gauge field variables.
The $\lambda_i$ are the eigenvalues of $\gamma_5 \cal M$.
One expects $\langle \Pi \rangle$ to be a good indicator
of the chiral limit at $\kappa\to\kappa_c(\beta)$, as some of
the $\lambda_i$ are expected to tend to zero.
It is worth noting that for periodic and time--antiperiodic
b.c. for the fermion fields the averages $\langle \Pi \rangle$
strictly coincide.

However, the numerical study of fermionic observables like
$\langle \Pi \rangle$ near the chiral limit does not reveal the
critical properties as expected from lowest order and finite lattice size
perturbation theory. This can be seen from the $\kappa$-dependence
of the pion norm numerically computed at low $\beta$-values within the
Coulomb phase \cite{HfMiMPNhSt} (see Fig. \ref{fig.1}).
Its behavior is very smooth and no sign of any critical behavior is observed.
The volume dependence of $\langle\Pi\rangle$ is rather weak,
and there is no significant difference between the quenched
and the dynamical case.

%=======================================================================
%   FIGURE 2.
%   Pion norm in the free fermion case.
%-----------------------------------------------------------------------
\begin{figure}[tbp]
\vspace{0.7cm}
\hspace*{0.5cm}
\epsfxsize=6.5cm\epsfysize=8.3cm\epsffile{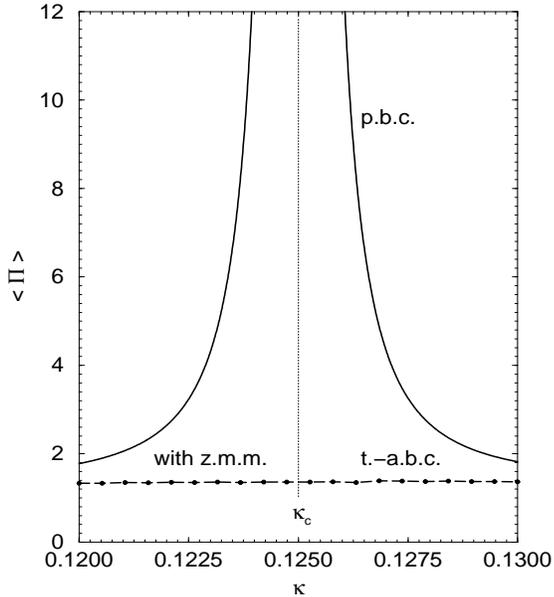}
\vspace{-0.1cm}
\caption{Pion norm in the free fermion case: without zero-momentum
modes and in the zero-momentum mode background, lattice size $6^4$,
periodic (p.b.c.), as well as time-antiperiodic (t-a.b.c.) boundary
conditions.}
\label{fig.2}
\end{figure}
%=======================================================================

It is interesting to compare these results for $\langle\Pi\rangle$
with the free fermion case given by

\begin{eqnarray}
\langle\Pi\rangle_0 &=& \frac{1}{V}\sum_p\left\{4\kappa^2
\sum_\mu \sin^2 \frac{2\pi p_\mu}{N_\mu} + \right.
\\
&& \left.
+ \Bigl(1-2\kappa\sum_\mu \cos\frac{2\pi p_\mu}{N_\mu}
\Bigr)^2 \right\}^{-1}\,,
\nonumber
\end{eqnarray}

\noi
where the $p_{\mu}, ~\mu=1,\cdots,4$ are integers except for 
time-antiperiodic b.c. causing $p_4$ to take half-integer values.
In Fig. \ref{fig.2} one can see the
$\kappa$--dependence of $\langle\Pi\rangle_0$ calculated on a 
symmetric lattice ($N_4 = N_s = 6$) for periodic and
time--antiperiodic b.c.
For periodic b.c. $\langle\Pi\rangle_0$ obviously
gets singular at $\kappa =1/8$, whereas
for time--antiperiodic b.c. the $\kappa$--dependence of
$\langle\Pi\rangle_0$ becomes smooth for symmetric
lattices. However, in the latter b.c. case 
$\langle\Pi\rangle_0$ develops a peak for strongly elongated lattices
($N_4\to\infty$ with $N_s$=fixed).

%=======================================================================
%   FIGURE 3.
%   Pion Norm in the Coulomb phase: comparison ZML vs. no gauge fixing.
%-----------------------------------------------------------------------
\begin{figure}[tbp]
\vspace{0.7cm}
\hspace*{0.5cm}
\epsfxsize=6.5cm\epsfysize=8.3cm\epsffile{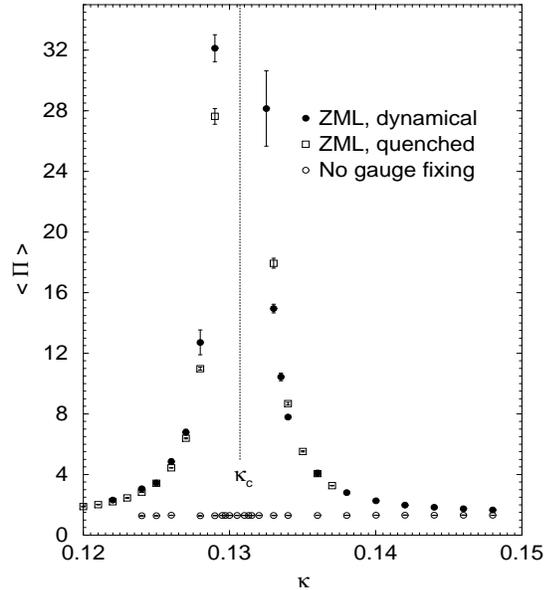}
\vspace{-0.1cm}
\caption{Pion norm as function of $\kappa$  
redefined with ZML gauge 
($\langle\Pi\rangle^{\prime}$) for full and quenched QED, as well as 
without any gauge fixing ($\langle\Pi\rangle$) for full QED; 
all data for $\beta=2.0$, lattice
size $4^4$, periodic b.c.'s.
}
\label{fig.3}
\end{figure}
%=======================================================================

We are going to demonstrate that this drastic difference between
$\langle\Pi\rangle$ (Fig. \ref{fig.1}) and $\langle\Pi\rangle_0$
(Fig. \ref{fig.2}) is due to constant or zero--momentum modes
of the gauge fields $z_\mu[U]$ as defined in eq.(\ref{zmm}).

For the free case ($\beta \to \infty$)
this can be easi\-ly demonstrated by integrating
$\langle\Pi\rangle_0$ over constant modes. Both for periodic
and time-antiperiodic b.c. it yields 
\begin{eqnarray}\label{eq8}
< \Pi > = \int\limits_{-\pi}^{\pi}\!\frac{{\rm d}^4\phi}{(2\pi)^4}%
\left\{4\kappa^2 \sum_\mu \sin^2 \phi_\mu + \right. \\
\left. + \Bigl(1-2\kappa\sum_\mu \cos\phi_\mu\Bigr)^2\right\}^{-1}\,.
\nonumber
\end{eqnarray}
The latter expression is completely smooth in $\kappa$ and agrees with
the former time-antiperiodic, free result for symmetric lattices
(see Fig. \ref{fig.2}).

Now let us consider the interacting case, i.e. for 
finite $\beta$-values. 
We redefine the pion norm such that zero-momentum modes become eliminated.
Each gauge field configuration generated in the
simulation is transformed subsequently by the ZML gauge procedure including
non-periodic gauge transformations in order to suppress
zero-momentum modes as described above. We average $\Pi$ with respect to
the ensemble of transformed fields. Note that the fermionic part of the 
action is not invariant under constant gauge transformations (\ref{npt}). 
Therefore, we get a new average   
$\langle\Pi\rangle^{\prime}$ which differs from $\langle\Pi\rangle$.

In the following we choose periodic boundary conditions, 
because we expect from the free case that they lead to a more
pronounced chiral behaviour than the time-antiperiodic ones.

In Figure \ref{fig.3} we show the dependence of the pion norm
$\langle\Pi\rangle$ and $\langle\Pi\rangle^{\prime}$ on $\kappa$.
One can see that for dynamical fermions (full circles) as well as
for quenched fermions (boxes) the redefined observable
$\langle\Pi\rangle^{\prime}$ has a sharp singularity near the point
$\kappa_c=0.1307(1)$ for $\beta=2.0$ \cite{BoMiMPPeZv}. In contrast,
the standard definition of the pion norm $\langle\Pi\rangle$ demonstrates
a completely smooth behavior (open circles). We checked these results for 
$\langle\Pi\rangle^{\prime}$ also on larger lattices. For 
$\kappa$ approaching $\kappa_c$  the same critical 
behaviour is observed, whereas very close to and slightly above $\kappa_c$ 
the influence of an increasing number of very small fermionic eigenmodes
leads to stronger fluctuations ('exceptional configurations').   
The dynamical and quenched results resemble each other. This can be 
interpreted such that the zero-momentum modes -- although removed from the
observable $\langle\Pi\rangle^{\prime}$ -- continue to dominate the 
fermionic determinant. 

What about other (maybe simpler) methods to get rid off the constant
modes (\ref{zmm}) of the gauge fields? We have also considered 
the Polyakov line gauge invented in \cite{BoMiMPPa} 
which transforms the spatially averaged Polyakov line
values into real numbers. We convinced ourselves that this non-periodic 
gauge -- without the necessity to employ the Lorentz gauge -- leads to the
same singular chiral behaviour of the pion norm as the ZML gauge. 

Our main conclusion is that zero--momentum modes play an important
r\^{o}le near the chiral limit of compact QED in the Coulomb phase, 
at least for finite lattices. They smooth out the critical chiral behaviour 
expected from lowest order perturbation theory for observables like 
the pion norm and the chiral condensate. In any case it is necessary
to take the zero-momentum modes properly into account. 

This research has been supported by the graduate college DFG-GK 271, 
by the RFRB grant 99-01-01230 and by the JINR Heisenberg-Landau program.

\end{document}